\documentstyle[12pt]{article}	
\newcommand{\bold}[1]{\mbox{\boldmath $#1$}}    
\newcommand{\gsim}{\raisebox{-0.5ex}{$\stackrel{>}{\sim}$}}

\newcommand{\k}{{\bf k}}                        
\newcommand{\r}{{\bold{r}}}                     
\newcommand{\ul}[1]{\underline{#1}}             
\newcommand{\MeV}{{\rm MeV}}                    
\newcommand{\fm}{{\rm fm}}                      
\newcommand{\mup}{\mu_\parallel}		
\newcommand{\mut}{\mu_\perp}			
\newcommand{\pphi}{{\bold{\phi}}}               
\newcommand{\ppsi}{{\bold{\psi}}}               
\newcommand{\del}{\partial}                     
\newcommand{\eps}{\epsilon}                     

\begin{document}
\baselineskip = 13.8pt plus 0.2pt minus 0.2pt   
    \lineskip = 13.8pt plus 0.2pt minus 0.2pt   

\noindent {\small {\sl RHIC Theory Workshop, 8-19 July 1996
\hfill LBNL-39332}}

\vspace{0.5in}
\noindent {\large {\bf Formation of DCCs with the linear sigma model}}
~\\[2ex]\noindent
J\o rgen Randrup\\[2ex]
Nuclear Science Division,
Lawrence Berkeley Laboratory,
Berkeley, California 94720
~\\[3ex]\noindent
A simple approximate treatment of statistical equilibrium
is developed within the linear $\sigma$ sigma model
by means of the Hartree factorization technique,
providing a simple means for sampling initial configurations
of the chiral field.
These are then subjected to pseudo-expansions
and their non-equilibrium relaxation towards the normal vacuum is studied.

~\\[3ex]\noindent
{\bf 1. INTRODUCTION}\\

\noindent
The possibility of forming disoriented chiral condensates
in high-energy hadron and heavy-ion collisions,
such as those anticipated at {\sl RHIC},
has generated considerable interest in the past few years.
The underlying assumption is that the collision produces an extended hot region
within which approximate chiral symmetry is temporarily restored.
The rapid disassembly of the system then induces a non-equilibrium relaxation
of the chiral field which may lead to the formation
of large coherent sources of correspondingly soft pions [1-5].
Since these disoriented chiral `domains'
have well-defined orientations in isospace
the associated pion multiplicity distributions display anomalous features.
The occurrence of the phenomenon may thus provide an observational opportunity
for testing our understanding of chiral symmetry. 
Recent reviews of the topic may be found in refs.\ [6-8].

In order to elucidate the conditions for the occurrence of {\sl DCC} phenomena
and the prospects for their experimental detection,
it is necessary to carry out dynamical simulations
of the non-equilibrium evolution experienced by the chiral field
as it relaxes from an initially very excited state,
in which chiral symmetry is approximately restored,
towards the normal vacuum in which the symmetry is significantly broken.
The most popular tool for such dynamical studies has been
the linear $\sigma$ model [9] in which the chiral degrees of freedom
are described by an $O(4)$ real classical field
with a simple non-linear self-interaction [10-20].
Even though this description is relatively simple,
ignoring all other degrees of freedom
(such as those represented by other mesonic species or individual partons),
it still presents a significant computational challenge.
It is therefore of practical interest to develop useful approximate methods
for solving the equation of motion for the chiral fields
and for understanding their complicated non-linear dynamics.

We first describe how it is possible to make a simple approximate treatment
of statistical equilibrium by employing the Hartree factorization technique.
We here largely follow the developments made in ref.\ [21].
Subsequently, following ref.\ [22],
we employ that framework for understanding the non-equilibrium relaxation
of chiral matter subjected to pseudo-expansions,
with particular emphasis on the conditions for achieving an amplification
of the low-energy pion modes.
A more comprehensive study of {\sl DCC} observables,
and the utility of the mean-field approximation for their calculation,
is being reported elsewhere [23].

\newpage\noindent
{\bf 2. HARTREE APPROXIMATION IN EQUILIBRIUM}\\

\noindent
Most dynamical studies of disoriented chiral condensates
have been based on the linear $\sigma$ model
in which the chiral degrees of freedom are described by the real $O(4)$ field
$\mbox{\boldmath $\phi$}=(\sigma,\mbox{\boldmath $\pi$})$
having the equation of motion
\begin{equation}
\label{EoM}
\left[\Box+\lambda(\phi^2-v^2)\right]\mbox{\boldmath $\phi$}\
=\ H\hat{\mbox{\boldmath $\sigma$}}\ ,\hspace{10mm}
\pphi(\r,t)\ =\ (\sigma(\r,t),\bold{\pi}(\r,t))\ .
\end{equation}
The three parameters in the model can be fixed by specifying
the pion decay constant, $f_\pi=92\ {\rm MeV}$, 
and the meson masses, $m_\pi=138\ {\rm MeV}$ and $m_\sigma=600\ {\rm MeV}$,
leading to the values
$\lambda=(m_\sigma^2-m_\pi^2)/2f_\pi^2=20.14$,
$v=[(m_\sigma^2-3m_\pi^2)/(m_\sigma^2-m_\pi^2)]^{1/2}f_\pi=86.71\ {\rm MeV}$,
and $H=m_\pi^2 f_\pi= (120.55\ {\rm MeV})^3$,
with $\hbar,c$=1 [21].

As is apparent from eq.\ (\ref{EoM}),
the vacuum configuration is aligned with the $\sigma$ direction,
$\mbox{\boldmath $\phi$}_{\rm vac}=(f_\pi,\mbox{\boldmath $0$})$,
and at low temperature the fluctuations
represent nearly free $\sigma$ and $\pi$ mesons.
In the other extreme, at temperatures well above $v$,
the field fluctuations are centered near zero
and approximate $O(4)$ symmetry prevails.

In the present discussion,
we limit the considerations to macroscopically uniform configurations
(chiral matter)
and therefore enclose the system in a box with periodic boundary conditions.
It is then possible to uniquely decompose the chiral field,
\begin{equation}
\mbox{\boldmath $\phi$}(\mbox{\boldmath $r$},t)\
=\ \underline{\mbox{\boldmath $\phi$}}(t)\
+\ \delta\mbox{\boldmath $\phi$}(\mbox{\boldmath $r$},t)\ ,\hspace{10mm}
\ul{\pphi}\ =\ <\pphi>\ .
\end{equation}
The first term, $\underline{\mbox{\boldmath $\phi$}}$,
is the spatial average of the chiral field
and may be identified with the order parameter,
while the fluctuations, $\delta\mbox{\boldmath $\phi$}(\mbox{\boldmath $r$})$,
represent elementary quasi-particle excitations relative to the constant field.

By taking the spatial average of the full equation of motion (\ref{EoM}),
it is possible to derive an equation of motion for the order parameter [24].
If we subsequently subtract that from (\ref{EoM})
and apply a Hartree-type factorization [16,21],
we obtain a corresponding equation for the field fluctuations.
The resulting equations of motion are of mean-field form,
\begin{eqnarray}\label{EoM0}
&~&\left[ \Box + \mu_0^2 \right] \ul{\pphi} = H\hat{\sigma}\ ,\hspace{10mm}
\mu_0^2\ =\ \lambda
[\phantom{3}\phi_0^2\ +\prec\delta\phi^2\succ
+\ 2\prec\delta\phi_\parallel^2\succ-\ v^2]\ ,\\
\label{EoMp}
&~&\left[ \Box + \mu_\parallel^2 \right] \delta\phi_\parallel = 0\
,\hspace{11mm} \mu_\parallel^2\ =\ \lambda
[{3}\phi_0^2\ +\prec\delta\phi^2\succ
+\ 2\prec\delta\phi_\parallel^2\succ -\ v^2]\ ,\\
\label{EoMt}
&~&\left[ \Box + \mu_\perp^2 \right] \delta\pphi_\perp= \mbox{\boldmath $0$}\
,\hspace{8mm} \mu_\perp^2\ =\ \lambda
[\phantom{3}\phi_0^2\ +\prec\delta\phi^2\succ
+\ 2\prec\delta\pphi_\perp^2\succ -\ v^2]\ .
\end{eqnarray}
Here $\delta\phi_\parallel=\delta\pphi\circ\hat{\ul{\pphi}}$
is the fluctuation along the order parameter
and $\delta\pphi_\perp$ is the fluctuation perpendicular to ${\ul{\pphi}}$.
Furthermore,
the spatial averages $<\cdot>$ have been replaced by the
corresponding thermal average $\prec\cdot\succ$,
evaluated at the given temperature $T$.
We note that the effective masses increase with
the magnitude of the order parameter $\phi_0$
as well as with the field fluctuations.
They are degenerate for $\phi_0$=0
and vanish at the temperature $T_c=\sqrt{2}v$.
Moreover, we always have $\mu_0^2\leq\mut^2\leq\mup^2$.
Since the quasi-particles are thus governed by a Klein-Gordon equation,
it is simple to obtain the thermal averages self-consistently,
\begin{eqnarray}
\prec\delta\phi_\parallel^2\succ &=& {1\over\Omega}
\sum_{\k\neq 0} {1\over\eps_k}{1\over{\rm e}^{\eps_k/T}-1}\ ,
\hspace{10mm} \eps_k^2\ =\ k^2\ +\ \mu_\parallel^2\ ,\\
\prec\delta\phi_\perp^2\succ &=& {1\over\Omega}
\sum_{\k\neq 0} {1\over\eps_k}{1\over{\rm e}^{\eps_k/T}-1}\ ,
\hspace{10mm} \eps_k^2\ =\ k^2\ +\ \mu_\perp^2\ .
\end{eqnarray}
The volume of the box is given by $\Omega$ and
$\k$ denotes the wave vector of the individual modes in the cavity.
The corresponding dispersion relations are indicated as well
and the resulting effective masses are shown in Fig.\ 1.

Eq.\ (\ref{EoM0}) was first derived in ref.\ [24]
and the Hartree treatment is in accordance with ref.\ [16].
Furthermore,
we note that the terms $<\delta\phi^2>$ in eqs.\ (\ref{EoM0}-\ref{EoMt})
are sums of the field fluctuations in each of the $N$=4 chiral directions and
thus constitute the leading-order fluctuation contribution in a $1/N$ expansion.
These are the `direct' terms that
have been included in a number of previous {\sl DCC} treatments or discussions
in terms of effective masses [10-11,16-19,25-27].
The additional fluctuation terms are the `exchange' terms
and each is twice the fluctuation along the particular chiral direction
considered (either parallel or perpendicular to the order parameter).
Although these terms are only of next-to-leading order in $1/N$,
they increase the fluctuation contributions by $2/N=50\%$
and their effect may thus be significant.
Recent analyses suggest that the mean-field treatment
with all the fluctuation terms included
is in fact a quite good approximation
for both the equilibrium properties [21]
and for the calculation of {\sl DCC} observables [23].

\begin{figure}[h]
\vspace{3mm}
\begin{minipage}{2.7in}
\vspace{60mm}\hspace{-16mm}
\includegraphics{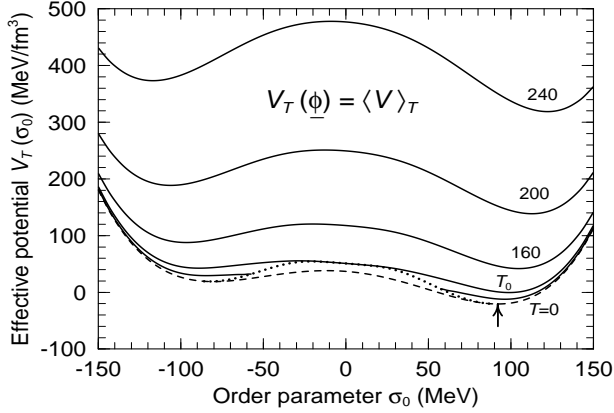}
\end{minipage}\hspace{25mm}
\begin{minipage}{2.7in}{\small
The effective potential energy density along the $\sigma$ axis
for a range of temperatures.
For $T$$<$$T_c$ the curve
starts at a minimum value between 0 and $v$;
these starting points are connected by the dotted curve,
while the dashed curve shows the bare potential $V_0$
for which $\delta\pphi\equiv\bold{0}$.
The potential for other orientations of the order parameter
can be obtained from
$V_T(\phi_0,\chi_0)$=$V_T(\phi_0,0)+H\phi_0(1-\cos\chi_0)$.
}\end{minipage}
\caption{Effective potential.}
\label{f:2}
\end{figure}

The statistical properties of chiral matter are most naturally derived
on the basis of the partition function,
\begin{equation}\label{Z}
{\cal Z}_T = \int{\cal D}[\ppsi,\pphi]
{\rm e}^{-{\Omega\over T}E[\ppsi,\pphi]} 
= \int d^4\ul{\ppsi} d^4\ul{\pphi}\ W_T(\ul{\ppsi},\ul{\pphi})\ ,\
W_T(\ul{\ppsi},\ul{\pphi}) \approx
{\rm e}^{-{\Omega\over T}(K_0 + V_T - TS_T)}\ ,
\end{equation}
where $K_0=\psi_0^2/2$ is the kinetic energy density of the order parameter
$\ul{\pphi}$.
The statistical weight $W_T$ gives the relative probability for finding
the system with a specified value of the order parameter $\ul{\pphi}$
and its time derivative $\ul{\ppsi}$.
Its simple approximate form contains
the effective potential energy density $V_T$,
which is shown in Fig.\ \ref{f:2},
and the entropy density $S_T$ associated with
the quasi-particle degrees of freedom for a given value of $\phi_0$.

\begin{figure}[h]
\vspace{3mm}
\begin{minipage}{2.7in}
\vspace{60mm}\hspace{-16mm}
\includegraphics{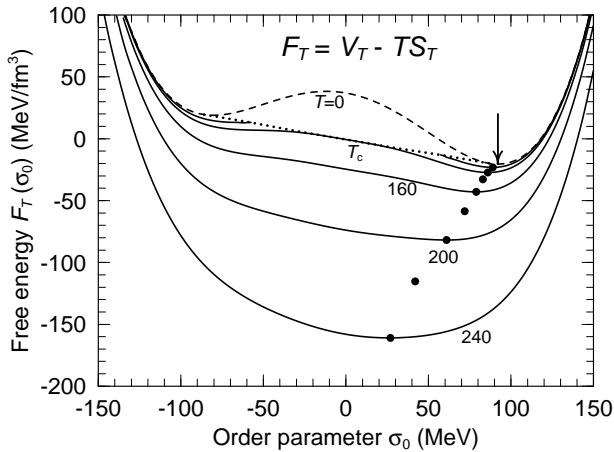}
\end{minipage}\hspace{25mm}
\begin{minipage}{2.7in}{\small
The free energy density along the $\sigma$ axis for a range of temperatures.
The solid curves show the results for a number of temperatures.
For $T$$<$$T_c$ the curve
starts at a minimum value of $\phi_0$
and these starting points are connected by the dotted curve,
while the dashed curve shows the result
obtained when the temperature is neglected.
The arrow points to the vacuum value $F_{T=0}$
and the location of the minima for finite temperatures
are indicated by the solid dots.
}\end{minipage}
\caption{Free energy.}
\label{f:3}
\end{figure}

The corresponding free energy density $F_T=V_T-TS_T$
governs the distribution of the order parameter $\ul{\pphi}$.
It depends only on the magnitude $\phi_0$ and the disorientation angle $\chi_0$
(the angle between $\ul{\pphi}$ and the $\sigma$ direction)
and it is easy to calculate in the mean-field approximation.
Figure \ref{f:3} shows its appearance along the $\sigma$ axis
where its minima are situated.
For high temperatures, the free energy has its minimum near the origin,
a reflection of the approximate $O(4)$ symmetry restoration,
and as $T$ is decreased,
the minimum moves smoothly outwards
and finally settles at the vacuum value $f_\pi$.

The location of the minimum in $F_T$ gives the 
most probable value of the order parameter.
For the relatively small volumes of relevance in collision experiments,
there are significant statistical fluctuations around the preferred value
and the corresponding thermal distribution is determined by
the statistical weight $W_T$ in (\ref{Z}).

~\newpage
\begin{figure}[t]
\begin{minipage}{2.7in}
\vspace{60mm}\hspace{-16mm}
\includegraphics{Fig1.ps}
\end{minipage}\hspace{25mm}
\begin{minipage}{2.7in}{\small
The effective masses $\mup$ and $\mut$
as functions of $\phi_0$ for a range of temperatures,
calculated in the thermodynamic limit, $L$$\to$$\infty$.
For temperatures above $T_c$,
the curves start at $\phi_0=0$ with degenerate values,
whereas below $T_c$
they only exist if $\phi_0$ is sufficiently large.
The corresponding starting points for $\mup$    
are connected by the dotted curve and,
since $\mup$ is then nearly independent of $T$,
only the curve for $T$=0 is shown.
The locations of the vacuum values $m_\sigma$, $m_\pi$ are also shown.
}\end{minipage}
\caption{Effective masses.}
\label{f:1}
\end{figure}

\newpage
\begin{figure}
\begin{minipage}{2.7in}
\vspace{80mm}\hspace{-16mm}
\includegraphics{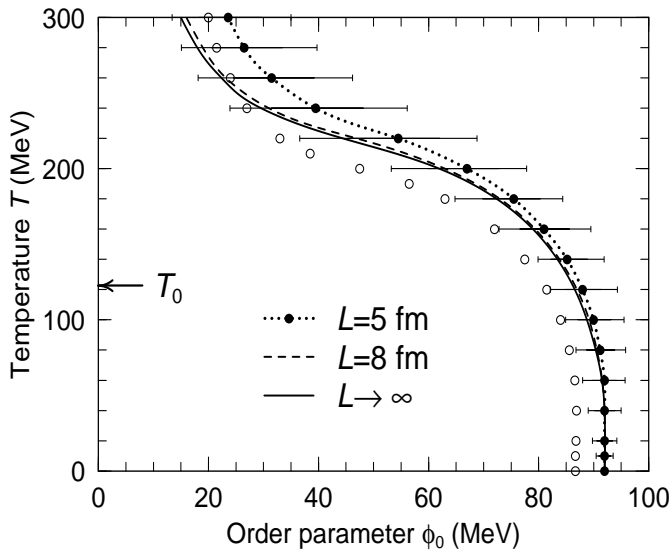}
\end{minipage}\hspace{25mm}
\begin{minipage}{2.7in}{\small
The most probable value of $\phi_0$, the magnitude of the order parameter,
in the standard case where $H>0$.
In the thermodynamics where $L\to\infty$ (solid curve)
$\phi_0$ is constrained to the value for which the free energy density
has its minimum (see Fig.\ \protect\ref{f:3}).
The bars show the full width at half maximum
of the thermal distribution of $\phi_0$ in the system with $L=5\ \fm$;
those for $L=8\ \fm$ are about half that size.
The open dots show the centroids for the idealized case having $H=0$,
for the box with $L=5\ \fm$;
the behavior is qalitatively similar even though the
nature of the phase transition changes.
}\end{minipage}
\caption{Temperature dependence of the order parameter.}
\label{f:4}
\end{figure}

The resulting behavior of the magnitude of the order parameter
is shown in Fig.\ \ref{f:4}.
As $T$ is increased from zero, the fluctuations grow steadily and
the equilibrium value of $\phi_0$ begins to decrease from $f_\pi$.
The most rapid change occurs at $T\approx\ 220\ \MeV$,
above which $\phi_0$ keeps decreasing at an ever slower rate.
Correspondingly,
$\mu_\perp$ increases monotonically with $T$
from its free value $m_\pi$ towards $\approx 1.6T$ for $T\gg T_c$,
while $\mu_\parallel$ first decreases,
then displays a minimum at $T\approx240\ {\rm MeV}$,
and finally becomes degenerate with $\mu_\perp$ [21].

~\\[3ex]\noindent
{\bf 3. SAMPLING OF THERMAL FIELD CONFIGURATIONS}\\

\noindent
The simple description resulting from the Hartree factorization
makes it possible to develop a convenient approximate manner
for sampling chiral field configurations
describing macroscopically uniform matter in thermal equilibrium [21].
Since this method is generally applicable,
it may be of broad interest and we therefore briefly summarize it here.

The first task is to sample the order parameter $(\ul{\ppsi},\ul{\pphi})$
on the basis of the statistical weight $W_T(\psi_0,\phi_0,\chi_0)$
given in eq.\ (\ref{Z}).
This quantity factorizes,
due to the additive form of the exponent.
The time derivative $\psi_0$ is then governed
by a four-dimensional normal distribution,
$P_\psi(\ul{\ppsi})\sim\exp(-\Omega\psi_0^2/2T)$.
Furthermore,
since the distribution of the magnitude $\phi_0$
can be pretabulated (ignoring at first the $H$ term),
the associated sampling task is computationally simple. 
Once $\phi_0$ has been picked,
the disorientation angle $\chi_0$ is easy to sample from
its conditional distribution,
$P_\chi(\chi_0)\sim\exp(-H\phi_0\cos\chi_0)$,
and the $O(3)$ direction $(\vartheta_0,\varphi_0)$
of $\ul{\bold{\pi}}$ is uniform on $4\pi$.

Once the magnitude of the order parameter is known,
the thermal quasiparticle distributions are fully determined
and the number of quanta in each elementary mode is readily sampled,
using $P(n_\k)\sim\exp(-n_\k\eps_k/T)$
for each of the four principal chiral directions.
Since the quasi-particle mass tensor is aligned with
the $O(4)$ direction of the order parameter, $(\chi_0,\vartheta_0,\varphi_0)$,
a subsequent $O(4)$ rotation of $\pphi(\r)$ and $\ppsi(\r)$
must then be performed in order to express the sampled field configuration
in the usual $(\sigma,\bold{\pi})$ reference system.

\newpage
\noindent
{\bf 4. EXPANSION DYNAMICS}\\

\noindent
We have shown above how thermal equilibrium can be treated approximately
in the mean-field approximation.
However,
it is expected that the early collision dynamics
causes the chiral field to be formed in a state of rapid expansion.
The subsequent evolution may then lead to
a supercooled configuration situated inside the unstable region,
thus effectively producing a ``quench''.
Several quenched scenarios have been considered [10-12,15-18]
but they were largely imposed by {\em fiat},
thereby reducing the predictive power of the dynamical calculations
- essentially any degree of magnification can be achieved
by suitable adjustment of the initial conditions.
The degree of arbitrariness can be reduced by examining
under which conditions a quench-like early scenario
may develop dynamically from various possible types of initial sceanrios.

Simple Bjorken-like pictures
can be invoked to emulate expansion in $D$ dimensions,
either longitudinal ($D$=1) [12,17-18],	
transverse ($D$=2), or isotropic ($D$=3) [13,27].	
We have studied such scenarios in an approximate manner
by augmenting the equation of motion (\ref{EoM})
with a Rayleigh dissipation term [22],
\begin{equation}\label{cool}
\left[\Box+\lambda(\phi^2-v^2)\right]\mbox{\boldmath $\phi$}\
-\ H\hat{\mbox{\boldmath $\sigma$}}\ =\ -{D\over t} \ppsi\ ,\hspace{10mm}
\ppsi\ =\ \del_t \pphi\ .
\end{equation}
The cooling term causes the field fluctuations to decay in the course of time
and the associated decrease of the energy density is given by
$\dot{E}=-(D/t)<\psi^2>$.
At sufficiently large times,
the quasiparticle number density decreases as $\sim t^{-D}$ as $t\to\infty$,
as is characteristic of an expansion in $D$ dimensions.
The time variable should now be reinterpreted as the elapsed proper time
in a comoving frame (starting at $t_0=1\ {\rm fm}/c$, usually).
The corresponding Lorentz transformation of the (scaled) spatial coordinates
is less essential for our present discussion
and has therefore been ignored.

\begin{figure}
\vspace{80mm}\hspace{-6mm}
\includegraphics{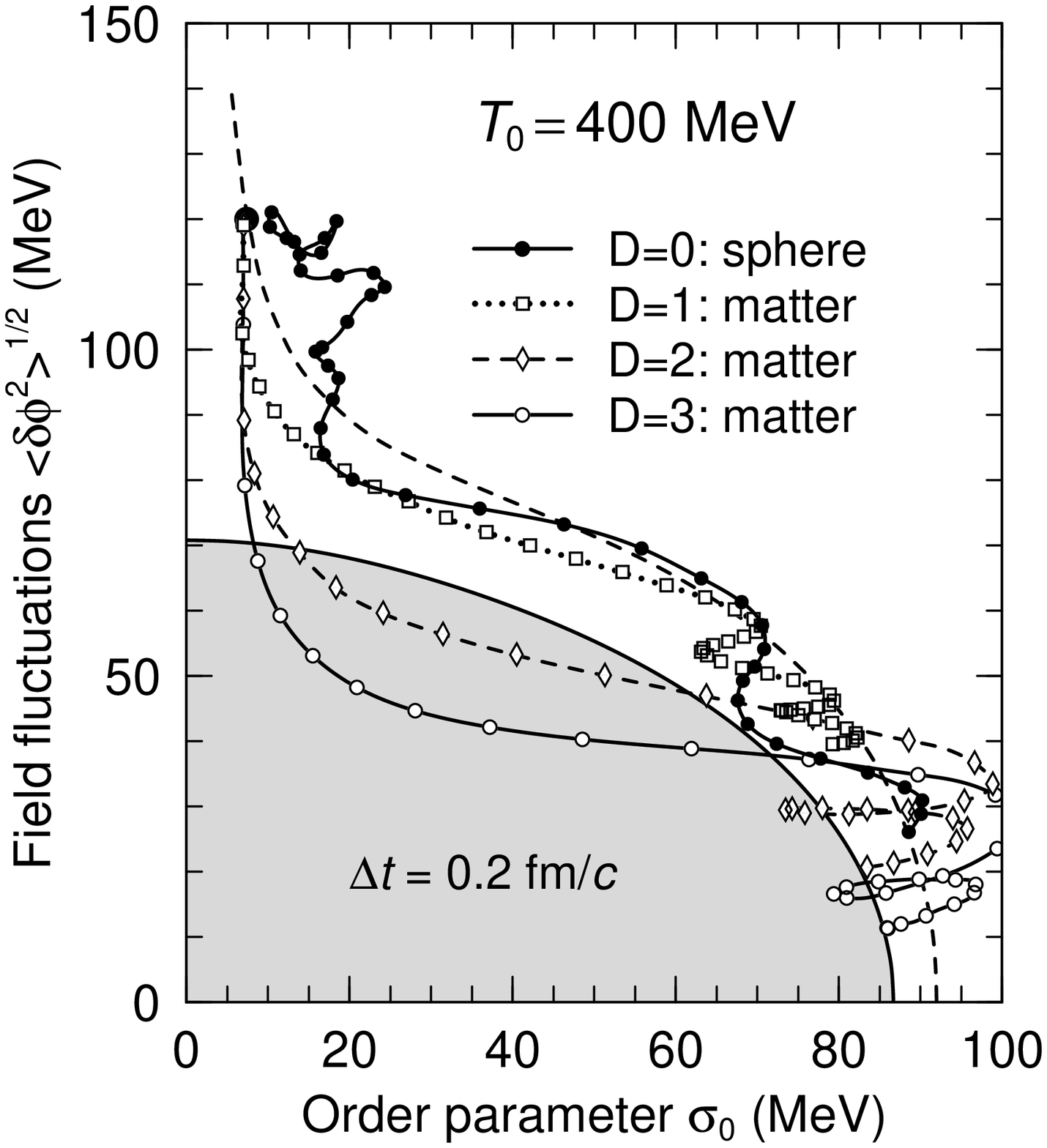}
\includegraphics{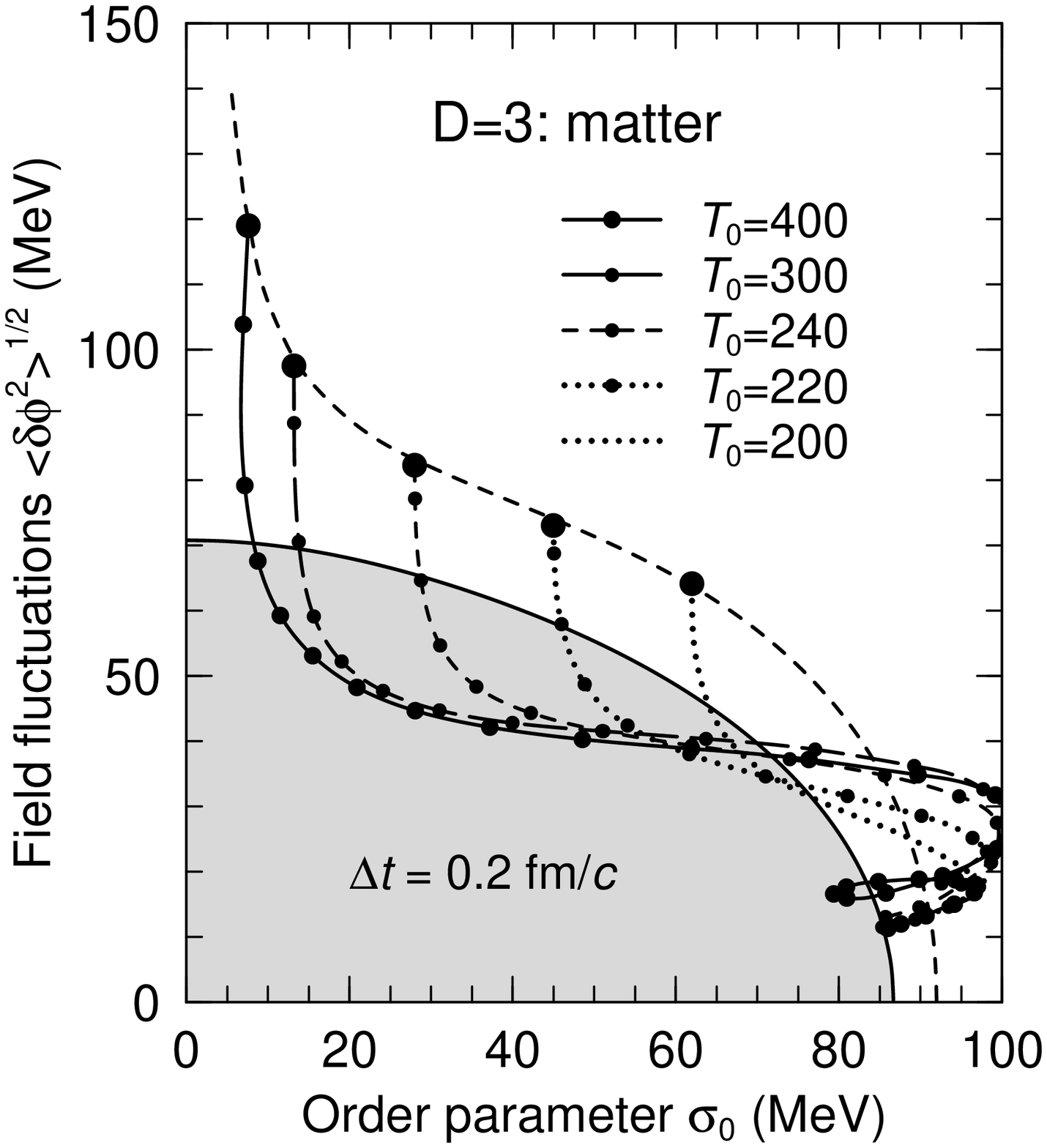}
{\small
The combined dynamical evolution
of the order parameter $\phi_0\approx\ <\sigma>$
and the fluctuations $<\delta\phi^2>^{1/2}$.
The dashed curve connects the equilibria
from $T$=0 to above 500 MeV
and the unstable region where $\mu_\perp^2<0$ is shown by the shaded region;
its border intersects the $\sigma$ axis at $f_\pi$
and extends up to $T_c/\sqrt{3}$ at $\phi_0$=0.
Each system has been prepared in thermal equilibrium at $T_0=400$ MeV,
using a periodic box (20 fm side length).
The irregular solid trajectory (labelled $D$=0)
was obtained by solving the standard eq.\ (\ref{EoM})
after applying a spherical Saxon-Woods modulation factor
(5 fm radius and 0.5 fm width) to the hot matter,
thereby producing a hot Ni-sizes sphere embedded in vacuum;
the field was sampled only in the interior ($r<$2.5).
The other three trajectories have been obtained
by solving the pseudo-expansion equation of motion (\ref{cool})
without applying a spatial modulation,
thus emulating uniform expansions in $D=1,2,3$ dimensions.
The marks along the trajectories are positioned
at time intervals of $\Delta t=0.2\ {\rm fm}/c$.
}
\caption{Dynamical trajectories.}
\label{f:5}
\end{figure}

Figures \ref{f:5}$a$-$b$ depict dynamical trajectories
for a variety of instructive scenarios.
In order to make a display that does not rely on any assumption
with regard to the degree of equilibrium,
we adopt the field dispersion as a measure of the degree of agitation;
it can be visualized as a model-independent replacement
of the temperature variable.
In Fig.\ \ref{f:5}$a$ is shown the dynamical trajectory of the central part
of a Ni-sized spherical source prepared at $T_0$=400 MeV
without any initial expansion.
The system keeps away from the unstable regime,
exhibiting an approximately adiabatic evolution.
This behavior is rather robust,
as it occurs for a wide range of initial temperatures
and for rod or slab geometries as well.
It thus appears that initially static field configurations
in local equilibrium do not develop any instabilities
during their subsequent expansion.

The other trajectories in Fig.\ \ref{f:5}$a$ illustrate the effect of
endowing the system with an initial expansion.
The effect increases with $D$,
since the dimensionality of the pseudo-expansion
effectively acts as the strength of the damping term.
The isotropic expansion leads to
a significant incursion into the unstable region,
while the longitudinal one is too slow for that
and appears to be closer to the self-generated near-adiabatic expansion.

The approximate equations (\ref{EoM0}-\ref{EoMt}) provide a convenient framework
for developing an understanding of the dynamics
generated by the pseudo-expansion (\ref{cool}).
Imagine that the system is initially created in thermal equilibrium
at a temperature $T_0$ well above $T_c$.
The field fluctuations are then sufficiently large to ensure $\mu^2$$>$0
in all three eqs.\ (\ref{EoM0}-\ref{EoMt}).
The system is expected to experience a cooling
resulting from expansion and radiation,
so the fluctuations decrease in the course of time.
This reduces $\mu^2$ which allows the order parameter to grow larger,
thus counteracting the decrease of the effective masses.
The resulting behavior of $\mu^2$ is then a delicate balance:
for slow cooling the induced growth of $\phi_0$ is approximately adiabatic
and the system relaxes towards the vacuum through metastable configurations;
however, if the fluctuations diminish rapidly
a compensating growth of the order parameter
can no longer occur quickly enough
and one or more of the effective masses may turn imaginary, $\mu^2$$<$0,
indicating that the system has entered an unstable regime
where some modes grow exponentiallly.

Figure \ref{f:5}$b$ shows trajectories for $D$=3
starting from various temperatures.
If the initial temperature is lower than 200 MeV or so,
the initial value of $\phi_0$ is already fairly large (over 60 MeV)
and the dynamical trajectories will miss the unstable region.
A wide range of higher temperatures lead into the unstable region,
provided the supercooling occurs sufficiently fast.
Ultimately, at very high temperatures (above those shown)
the system will again stay stable throughout,
because it takes so long to reduce the fluctuations down to critical size
that the order parameter has meanwhile had time to start its growth.

\newpage
\begin{figure}
\begin{minipage}{2.7in}
\vspace{80mm}\hspace{-16mm}
\includegraphics{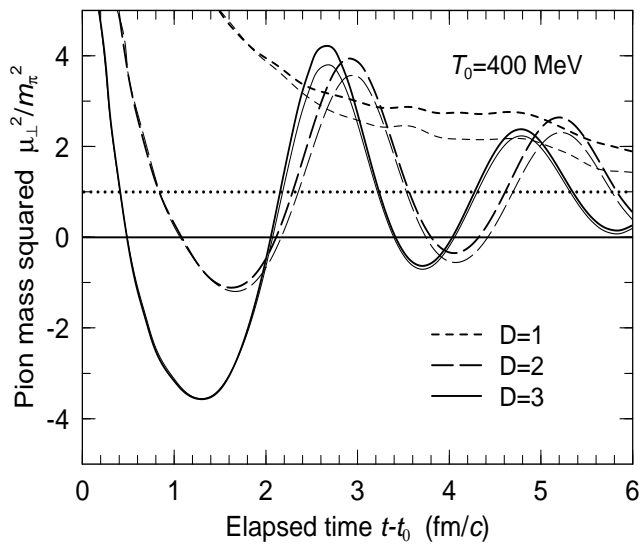}
\end{minipage}\hspace{25mm}
\begin{minipage}{2.7in}{\small
The time evolution of $\mu_\perp^2$ for pseudo-expansions (\ref{cool})
with various values of the dimensionality $D$,
starting from thermal equilibrium at $T_0$=400 MeV (heavy curves),
and the corresponding evolution of $\mu_0^2$ (thin curves).
The evolution is started at $t_0=1\ \fm/c$,
as is commonly done, and 
all curves approach the free pion mass $m_\pi$ (dotted line) for $t\to\infty$.
Whenever $\mu_\perp^2<0$
those transverse modes having $k^2<-\mu_\perp^2$ are amplified
and the resulting maximum degree of amplification is illustrated
in Table 1 for a range of initial temperatures.
}\end{minipage}
\caption{Time evolution of the pion mass.}
\label{f:7}
\end{figure}

In order to quantify the analysis,
it is useful to consider the time evolution of the effective masses.
Since $\mu_\parallel^2>\mu_\perp^2$
we concentrate on the latter which is illustrated in Fig.\ \ref{f:7}.
It is noteworthy that $\mu_0\approx\mu_\perp$ throughout the evolution,
implying that the amplification of the lowest pionic modes
is practically identical to that of the order parameter itself.
This simple feature makes it an easier task to analyze
more complicated scenarios.
It is convenient to express the resulting enhancement of a mode
in terms of its amplification coefficient [28],
\begin{equation}\label{G0}
G^\pi_k\ \equiv\	
\int_{\omega_k^2<0}dt\ \sqrt{-\omega_k(t)^2}\ ,\hspace{10mm}
\omega_k^2=k^2+\mu_\perp(t)^2\ ,
\end{equation}
which expresses approximately the factor by which the amplitude of a pionic mode
has been magnified due to incursion(s) into the unstable regime.
An upper bound on the magnification
is provided by the coefficient for $k$=0, shown in Table \ref{table}.
The purely longitudinal expansions largely miss the unstable region,
while significant magnification occurs for
the transverse and isotropic expansions,
amounting to over a factor of ten in the most favorable cases.

\vfill
\begin{table}[h]
\caption{Amplification coefficients/correlation lengths.}
\label{table}
\vspace{2mm}
\begin{minipage}{3.0in}
\begin{tabular}{|c|ccc|}
\hline	
$T_0$ (MeV)	& $D=1$		& $D=2$		& $D=3$		\\
\hline	
200		& 0.00	/1.4	& 0.02	/1.8	& 0.11	/2.0	\\
220		& 0.00	/1.3	& 0.50	/1.9	& 0.55	/2.5	\\
240		& 0.01	/1.3	& 1.20	/2.0	& 1.19	/2.7	\\
300		& 0.00	/0.9	& 1.84	/1.7	& 2.06	/2.7	\\
400		& 0.00	/0.6	& 1.67	/1.3	& 2.49	/2.1	\\
500		& 0.00	/0.5	& 1.31	/1.1	& 2.61	/1.6	\\
\hline	
\end{tabular}	
\end{minipage}\hspace{0.5in}
\begin{minipage}{2.9in}{\small
The maximum enhancement factor, $G^\pi_{k=0}$,
in macroscopically uniform chiral matter
and the resulting {\sl FWHM} of the pion correlation function
$C_\pi(\r_{12})=\ <\delta\mbox{\boldmath $\pi$}(\r_1)\cdot
\delta\mbox{\boldmath $\pi$}(\r_2)>$
for pions emerging after pseudo-expansions (\ref{cool})
starting from thermal equilibrium at the temperature $T_0$ and using $D=1,2,3$.
}\end{minipage}
\end{table}

\newpage
\begin{figure}
\begin{minipage}{2.7in}
\vspace{80mm}\hspace{-16mm}
\includegraphics{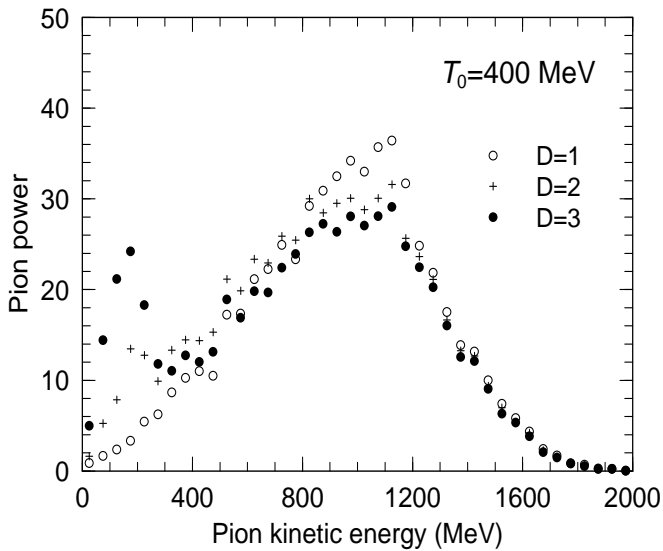}
\end{minipage}\hspace{25mm}
\begin{minipage}{2.7in}{\small
The relative power spectrum of the pions,
$\sim\omega_k^2\pi_k^2$,
where $\mbox{\boldmath $\pi$}_k$ is the Fourier amplitude of the pion field,
plotted as a function of the pion kinetic energy $\omega_k$-$m_\pi$.
The extraction is made at large times
when the asymptotic scenario of free evolution has been reached.
The plots are based on samples of 20 field configurations
prepared at $T_0$=400 MeV
and subjected to idealized expansions
with either $D$=1 (open dots), $D$=2 (crosses), or $D$=3 (solid dots).
The irregularities are primarily due to the shell structure
in the level density of the cube.
}\end{minipage}
\caption{Pion power spectrum.}
\label{f:8}
\end{figure}


Figure \ref{f:8} gives an impression of the net effect
on the power spectrum of the emerging free pions.
As expected, the transient instabilities present for $D\gsim2$
lead to significant enhancements of the power carried off by soft pions.
The effect amounts to about an order of magnitude for $D$=3
(relative to the smooth spectrum obtained for $D$=1),
in accordance with the amplification coefficients given in Table \ref{table}.
Although these results were calculated for idealized expansion scenarios,
they do support the suggestion that such enhancements
may provide an observable signal of {\sl DCC} formation [7].

~\\[3ex]
\noindent
{\bf 5. CONCLUDING REMARKS}\\

\noindent
We have discussed some of the key features associated with the formation
of disoriented chiral condensates in high-energy collisions,
such as those planned at {\sl RHIC}.
By application of the Hartree factorization technique,
it is possible to develop a simple mean-field treatment
which in turn leads to an efficient method for sampling thermal field
configurations.
Moreover, the mean-field approximation provides a useful conceptual framework
for understanding the non-equilibrium dynamics of the chiral field
as it relaxes from its initial very excited state towards the normal vacuum.
By augmenting the full equation of motion with a cooling term
it is possible to emulate chiral matter in uniform Bjorken-like expansion.

With this method, we have studied
the conditions for amplification of the soft pionic modes,
an important element in the observation of {\sl DCC}s.
Our analysis shows that the occurrence of instabilities,
and the associated amplification of pionic modes,
depends sensitively on the cooling rate,
which in turn is intimately related to the character of the expansion.
Our idealized scenario for $D$=3
corresponds closely to the isotropic expansion
considered in refs.\ [13,27]
and our results corroborate the conclusion in [13]
that such a scenario leads to amplification.
Furthermore,
our analysis suggest that a longitudinal expansion alone
is insufficient to cause a quench,
if the initial fluctuations are of thermal magnitude.
This is consistent with what was found in refs.\ [12,18]
for effectively one-dimensional expansions.

This qualitative sensitivity to the collision dynamics
highlights the importance of employing realistic initial conditions
for the dynamical simulations of {\sl DCC} formation.
Ultimately,
the appropriate initial field configurations
must be calculated on the basis of the early partonic evolution,
a task which is thus crucial for our ability to assess the prospects
of forming disoriented chiral condensates in high-energy collisions.\\

\noindent
This work was supported in part by the Director,
Office of Energy Research,
Office of High Energy and Nuclear Physics,
Nuclear Physics Division of the U.S. Department of Energy
under Contract No.\ DE-AC03-76SF00098.

~\\[3ex]\noindent
{\bf REFERENCES}\\[-2ex]

\noindent \phantom{1}1.
A.A. Anselm, Phys. Lett. B217, 169 (1989).\\[-2ex]

\noindent \phantom{1}2.
A.A. Anselm and M.G. Ryskin, Phys. Lett. B266 (1991) 482\\[-2ex]

\noindent \phantom{1}3.
J.-P. Blaizot and A. Krzywicki, Phys. Rev. D46 (1992) 246 \\[-2ex]

\noindent \phantom{1}4.
K. Rajagopal and F. Wilczek, Nucl. Phys. B399 (1993) 395\\[-2ex]

\noindent \phantom{1}5.
J.D. Bjorken, K.L. Kowalski, and C.C. Taylor,
Report SLAC-PUB-6109 (1993)\\[-2ex]

\noindent \phantom{1}6.
K. Rajagopal, in {\sl Quark-Gluon Plasma 2},
ed.\ R. Hwa, World Scientific (1995)\\[-2ex]

\noindent \phantom{1}7.
S. Gavin, Nucl. Phys. A590 (1995) 163c\\[-2ex]

\noindent \phantom{1}8.
J.-P. Blaizot and A. Krzywicki, Acta Physica Polonica B;
hep-ph/9606263 (1996)\\[-2ex]

\noindent \phantom{1}9.
M. Gell-Mann and M. Levy, Nuovo Cimento 16 (1960) 705\\[-2ex]

\noindent 10.
K. Rajagopal and F. Wilczek, Nucl. Phys. B404 (1993) 577\\[-2ex]

\noindent 11.
S. Gavin, A. Gocksch, and R.D. Pisarski, Phys. Rev. Lett. 72 (1994) 2143\\[-2ex]

\noindent 12.
Z. Huang and X.-N. Wang, Phys. Rev. D49 (1994) 4335\\[-2ex]

\noindent 13.
S. Gavin and B. M\"uller, Phys. Lett. B329 (1994) 486\\[-2ex]

\noindent 14.
J.-P. Blaizot and A. Krzywicki, Phys. Rev. D50 (1994) 442\\[-2ex]

\noindent 15.
M. Asakawa, Z. Huang, and X.-N. Wang, Phys. Rev. Lett. 74 (1995) 3126\\[-2ex]

\noindent 16.
D. Boyanovsky, H.J. de Vega, and R. Holman, Phys. Rev. D51 (1995) 734\\[-2ex]

\noindent 17.
F. Cooper, Y. Kluger, E. Mottola, and J.P. Paz,
Phys. Rev. D51 (1995) 2377\\[-2ex]

\noindent 18.
Y.~Kluger, F.~Cooper, E.~Mottola, J.P.~Paz, A.~Kovner,
Nucl.~Phys.~A590 (1995) 581c\\[-2ex]

\noindent 19.
S. Mr\'owczy\'nski and B. M\"uller, Phys. Lett. 363B (1995) 1\\[-2ex]

\noindent 20.
L.P. Csernai and I.N. Mishustin, Phys. Rev. Lett. 74 (1995) 5005\\[-2ex]

\noindent 21.
J. Randrup, Phys. Rev. D (submitted); Report LBL-38125 (1996)\\[-2ex]

\noindent 22.
J. Randrup, Phys. Rev. Lett. 77 (1996) 1226\\[-2ex]

\noindent 23.
J. Randrup, Nucl. Phys. A (submitted); Report LBNL-39328 (1996)\\[-2ex]

\noindent 24.
G. Baym and G. Grinstein, Phys. Rev. D15 (1977) 2897\\[-2ex]

\noindent 25.
D. Boyanovsky and D.L. Lee, Phys. Rev. D48 (1993) 800\\[-2ex]

\noindent 26.
R.D. Amado and I.I. Kogan, Phys. Rev. D51 (1995) 190\\[-2ex]

\noindent 27.
M.A. Lampert, J.F. Dawson, and F. Cooper, Preprint hep-th/9603068 (1996)\\[-2ex]

\noindent 28.
H. Heiselberg, C.J. Pethick, and D.G. Ravenhall, Phys. Rev. Lett. 61 (1988) 818

			\end{document}